\documentclass[aps,twocolumn,showpacs,preprintnumbers,amsmath,amssymb,superscriptaddress]{revtex4-2}

\usepackage{hyperref}
\usepackage{amssymb}
\usepackage{color,dcolumn}
\usepackage{booktabs}
\usepackage{siunitx,physics}

\def \Prague{Charles University, Faculty of Mathematics and Physics, Institute of Physics, Ke Karlovu 5, Prague 2, CZ-121~16, Czech Republic}
\def \MU{Masaryk University, Department of Condensed Matter Physics, Kotlářská 267/2, Brno, CZ-611~37, Czech Republic}
\def \VUT{CEITEC - Central European Institute of Technology, Brno University of Technology, Purkyňova 123, Brno, CZ-612~00, Czech Republic}

\usepackage{epsfig}
\usepackage{amsmath}

\begin{document}
\title{Unveiling Davydov-Split Excitons in a Template-Engineered Molecular-Graphene Heterostructure}
\author{J. \surname{Kunc}} \affiliation{\Prague}
\email{jan.kunc@matfyz.cuni.cz}
\author{B. \surname{Morzhuk}} \affiliation{\Prague}
\author{V. \surname{Stará}} \affiliation{\VUT}
\author{D. \surname{Varshney}} \affiliation{\MU}
\author{M. \surname{Shestopalov}} \affiliation{\Prague}
\author{K. \surname{Matějka}} \affiliation{\Prague}
\author{M. \surname{Rejhon}} \affiliation{\Prague}
\author{J. \surname{Novák}} \affiliation{\MU}
\author{J. \surname{Čechal}} \affiliation{\VUT}

\date{\today}

\begin{abstract}
The realization of high-fidelity organic-inorganic quantum emulators is frequently hindered by the interfacial imperfections introduced during device fabrication. Here, we demonstrate a robust nanofabrication protocol that restores the atomic-scale purity of epitaxial graphene on SiC to UHV-equivalent levels, as confirmed by Low-Energy Electron Diffraction, and Microscopy. This pristine interface enables the emergence of macroscopic excitonic coherence in epitaxial overlayers of 2,3,6,7,10,11-hexamethoxytriphenylene (HMTP), a model molecular system characterized by intense electron-phonon coupling. Through a combination of high-sensitivity Fourier Transform Photo-current Spectroscopy, photoluminescence, and dynamic Raman mapping, we resolve a complex vibronic manifold governed by Davydov splitting. We show that the $P6_3/m$ crystalline symmetry of the HMTP overlayer lifts the degeneracy of the HOMO-LUMO transition, creating discrete bright and dark excitonic branches. Using an analytical tight-binding model parameterized by ARPES-derived intermolecular coupling and Raman vibrational modes validated by molecular dynamics simulations, we quantify the polarization energy, the Huang-Rhys factor, and Herzberg-Teller corrections to the Franck-Condon model. Our results reveal that the dark-state branch dominates the radiative channel, following a polaron-mediated relaxation pathway consistent with Kasha's rule. By reconciling macroscopic device architecture with UHV-level surface science, this work establishes a scalable platform for the study of dark-exciton dynamics and the development of solid-state molecular quantum memories.
\end{abstract}


\maketitle
\section{Introduction}
The simulation of open quantum systems, where electronic degrees of freedom are coupled to a complex bosonic environment, remains a frontier challenge in both condensed matter physics and quantum information science~\cite{Ref1LiuWuJCP150-105102-2019,Ref2WestJPCC114-10580-2010,Ref3WestJPCB5157-2010}. Central to this challenge is the Holstein Hamiltonian~\cite{Ref4KistlerJPCB116-77-2012,FetherolfPRX2020}, which describes the non-adiabatic dynamics of excitons and polarons—phenomena that are often computationally intractable in the strong-coupling regime~\cite{Ref5GiavazziJCTC436-2023}. While synthetic quantum simulators such as trapped ions~\cite{MezzacapoPhysRevLett2012trappedIons,GormanPhysRevX.8.011038} and superconducting circuits~\cite{HouckNatPhys2012} attempt to approximate these models, the realization of a physical, atomically well-defined platform remains elusive. Although graphene-based architectures have been proposed as highly tunable environments for molecular electronics~\cite{RissCrommieACSNano8-5395-2014}, achieving the necessary structural order to emulate complex vibronic dynamics has proven difficult. Here, we demonstrate that self-assembled layers of 2,3,6,7,10,11-hexamethoxytriphenylene (HMTP) on epitaxial graphene provide a pristine, structurally ordered emulator for these complex vibronic systems. The observation of intrinsic molecular dynamics is historically hindered by the pervasive issue of polymer contamination during the nanofabrication of graphene-based devices~\cite{HussHansenLangmuir2020,Matsumae2016-cleaning,YagerAPL106-063503-2015,TyagiColettiNanoscale14-2167-2022}. Traditional lithographic workflows necessitate multiple spin-coating cycles, leaving behind cross-linked resist residues~\cite{PirkleAPL99-122108-2011} that can quench excitonic states and obscure surface-sensitive probes. Rather than relying on simple self-assembly~\cite{CrommieACSNano7-6123-2013} or specialized resist-less masking~\cite{Wei2020}, we overcome this bottleneck through a synergistic fabrication strategy: a streamlined two-step electron-beam lithography process that minimizes resist exposure, followed by a specialized wet-chemical cleaning procedure utilizing dioxolane. The efficacy of this approach is evidenced by the emergence of sharp diffraction spots in Low-Energy Electron Diffraction (LEED) and high-contrast Low-Energy Electron Microscopy (LEEM)—demonstrating a level of interfacial purity post-fabrication that, to our knowledge, has not been previously achieved for functionalized graphene devices.

Parallel to achieving interfacial purity, the extreme photosensitivity of organic semiconductors~\cite{AndersonAPB120-2015-Ramandifficulty,Gottfried2013} necessitated a dynamic Raman acquisition protocol to prevent photodegradation~\cite{SurovtsevaJAS2010}. By employing sub-milliwatt excitation and rapid lateral scanning, we resolved previously undocumented vibrational modes essential for parameterizing the system's vibronic coupling constants. Crucially, the resulting pristine interface allows the epitaxial graphene lattice to serve as a critical structural template that dictates the macroscopic coherence of the molecular overlayer~\cite{CrommieACSNano7-6123-2013}. While growth on metallic, or dielectric substrates~\cite{WittePSSA2008} typically exhibits randomized azimuthal distribution, our LEED analysis confirms that HMTP growth on graphene results in an exceptionally well-ordered phase~\cite{DevanshuTBS,DevanshuArxiv} locked to two discrete, symmetry-equivalent orientations. This template-induced alignment is the prerequisite for resolving the fundamental vibronic structure~\cite{HernandoPRL97-216403-2006,Ref6TavazziJPCC118-8588-2014,Nakatsu1977} and transition dipole orientations of the molecular ensemble. Specifically, this crystalline order reveals the signature of Davydov splitting—a symmetry-breaking intermolecular exchange interaction~\cite{BeljonneJCP112-2000} that generates an upper 'bright' and a lower 'dark' excitonic branch. Leveraging the spectral resolution of this pristine interface, we synthesize results from Fourier Transform Photocurrent Spectroscopy, Angle-Resolved Photoemission Spectroscopy (ARPES), and molecular dynamics to extract the fundamental constants of the Holstein Hamiltonian with unprecedented precision. We quantify the Davydov splitting strength ($J$), Huang-Rhys factors ($S$), and the relative contributions of Franck-Condon versus Herzberg-Teller coupling mechanisms. Our findings demonstrate that the HMTP-graphene heterostructure operates in a regime where intermolecular electronic coupling is comparable to the vibrational relaxation energy, providing a rigorous testbed for non-perturbative theoretical models~\cite{Ref4KistlerJPCB116-77-2012,Ref5GiavazziJCTC436-2023}. By mapping the specific phonon modes and spectral density of the bosonic environment, we establish a physical benchmark for quantum simulators to probe the interplay between dissipation, symmetry, and electronic coherence in an atomically well-defined environment.

\section{Results}
\begin{figure*}[t!]
\centering
\includegraphics[width=18cm]{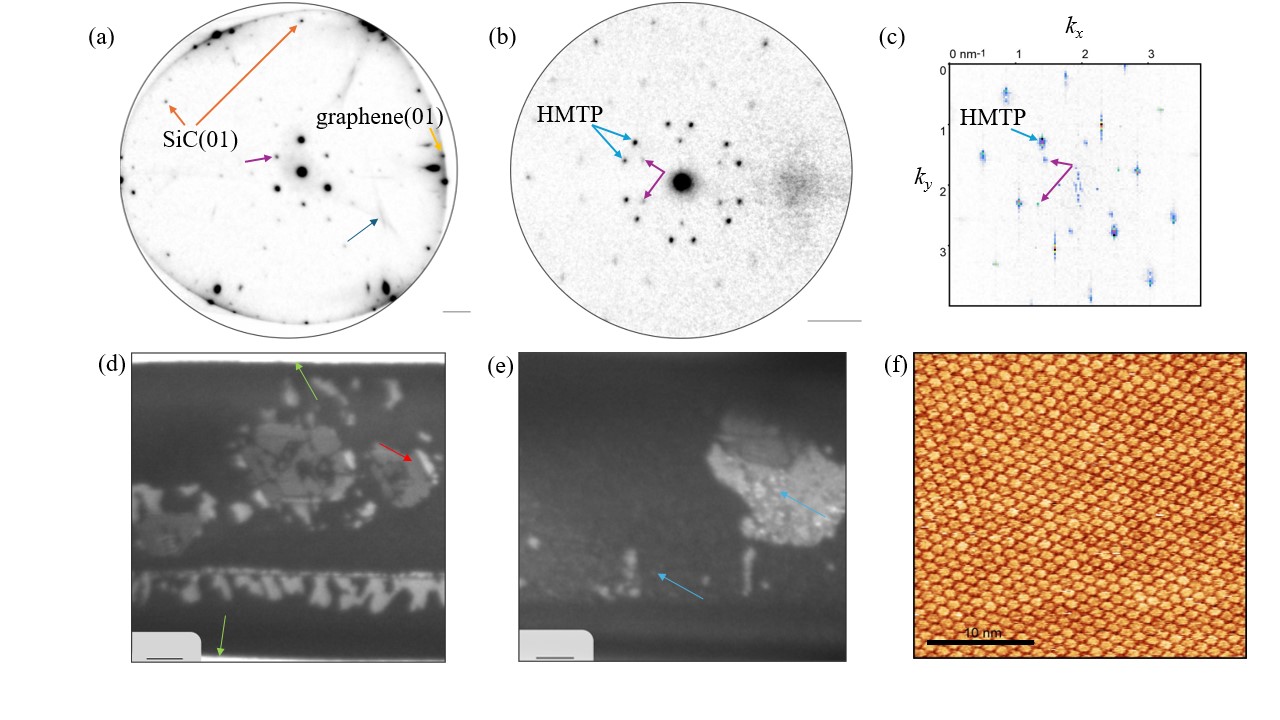}
\caption{Structural characterization of HMTP submonolayers. LEED pattern of the graphene interdigitated device following the post-nanofabrication cleaning procedure (a) before, and (b) after HMTP submonolayer growth; the sharp diffraction spots in (a) confirm the removal of polymer residues. (c) Numerical Fourier Transform (FFT) of the STM image in (f), showing reciprocal space points consistent with the LEED symmetry in (b). (d) LEEM image of a 3 $\mu$m wide graphene electrode before and (e) after HMTP deposition; scale bars represent 0.5~$\mathrm{\mu m}$. (f) High-resolution STM image of the HMTP layer on an unpatterned graphene substrate, revealing the molecular self-assembly.}
\label{FigLEED}
\end{figure*} 
\begin{figure*}[t!]
\centering
\includegraphics[width=18cm]{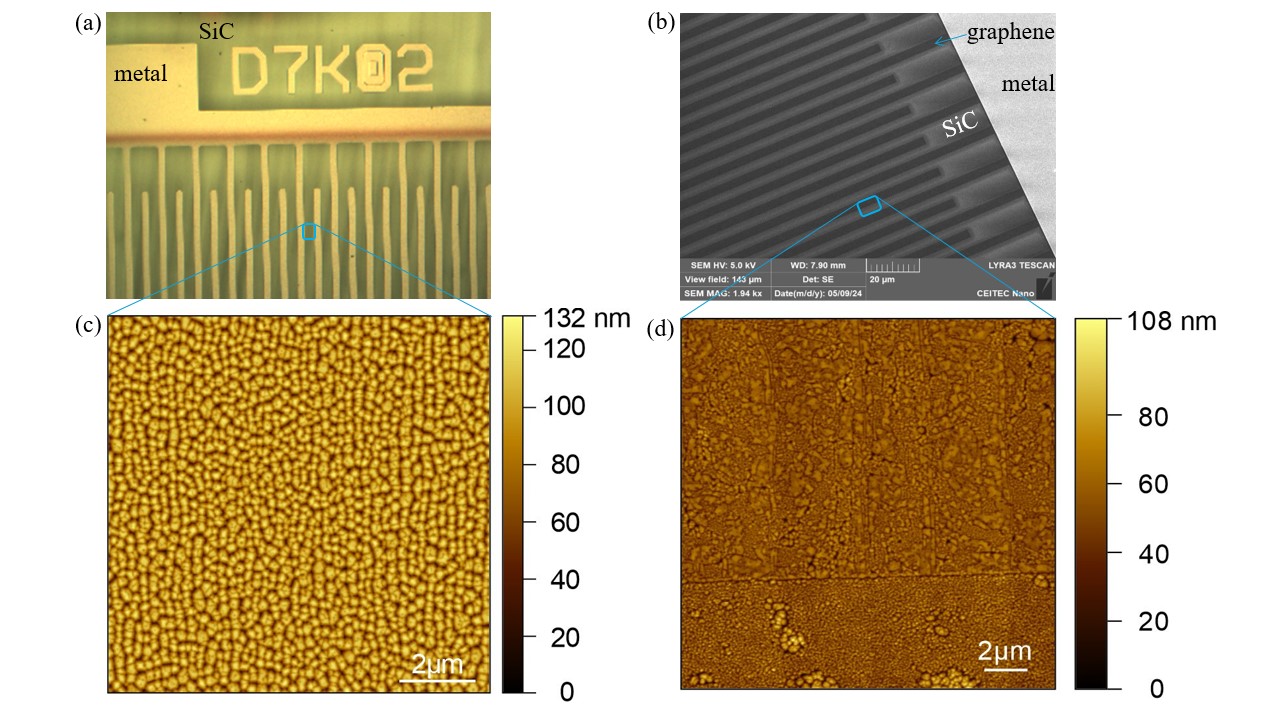}
\caption{
Morphological characterization of HMTP-based devices. (a) Optical micrograph of a Fourier transform photocurrent device featuring the prefabricated metal interdigitated electrode array. (b) SEM image of interdigitated contacts patterned from epitaxial graphene on SiC. (c) AFM topography of a 50 nm thick HMTP film deposited onto the metal-electrode platform shown in (a). (d) AFM topography of the 50 nm HMTP film at the interface between the graphene contact (upper region) and the bare SiC substrate (lower region).}
\label{FigDevice}
\end{figure*} 
\subsection{Interfacial Engineering and Device Fabrication}
\label{sec:Fabrication}
The realization of intrinsic exciton dynamics requires an interface free from the cross-linked polymer residues typical of standard lithographic processing. To this end, we developed a streamlined fabrication protocol that maximizes process reliability by eliminating the chemical complexity of multi-developer workflows. Our approach utilizes a four-layer resist stack—comprising Poly(methyl methacrylate) (PMMA) 200$^{\circ}$C / methyl methacrylate (MMA) 200$^{\circ}$C / MMA 180$^{\circ}$C and a top layer of Electra AR-PC 5090 for charge dissipation—to facilitate edge-contacted graphene devices in a two-step lithographic process, see Appendix~\ref{sec:Fab} Fig.~\ref{FigEdgeContactFab} and Tab.~\ref{tab:fabrication}. While previous two-step architectures have utilized mixed PMMA and CSAR resist stacks, these require a sequential development process in $o$-xylene followed by an isopropanol (IPA)/water mixture~\cite{ShettyACSAppl2023}. In contrast, our all-methacrylate stack is optimized for a single-step development in AR-P 600-56. By carefully tuning the baking temperatures of the MMA layers to control the profile of the undercut, we achieve high-fidelity lift-off and consistent edge-contact formation without the need for multiple solvent systems. Process optimization revealed that a restricted development time of 30 s and a precise oxygen plasma etch of 11–15 s were critical to maintaining the structural integrity of the graphene lattice while ensuring low-resistance ohmic junctions. Transfer Length Method (TLM) measurements confirm the efficacy of this process. After accounting for local graphene inhomogeneities and minor contact variations, we extract a contact resistance ($R_c$) of approximately $1000$–$1200\ \Omega \cdot \mu\text{m}$. Notably, our analysis indicates that this simplified development cycle yields a specific contact resistance competitive with more chemically complex triple-resist methods~\cite{ShettyACSAppl2023}. This suggests that the reduction in chemical steps is yet a viable route that enhances the structural quality of the graphene interface, providing the necessary foundation for the subsequent surface cleaning, and molecular deposition.
\begin{figure*}[t!]
\centering
\includegraphics[width=18cm]{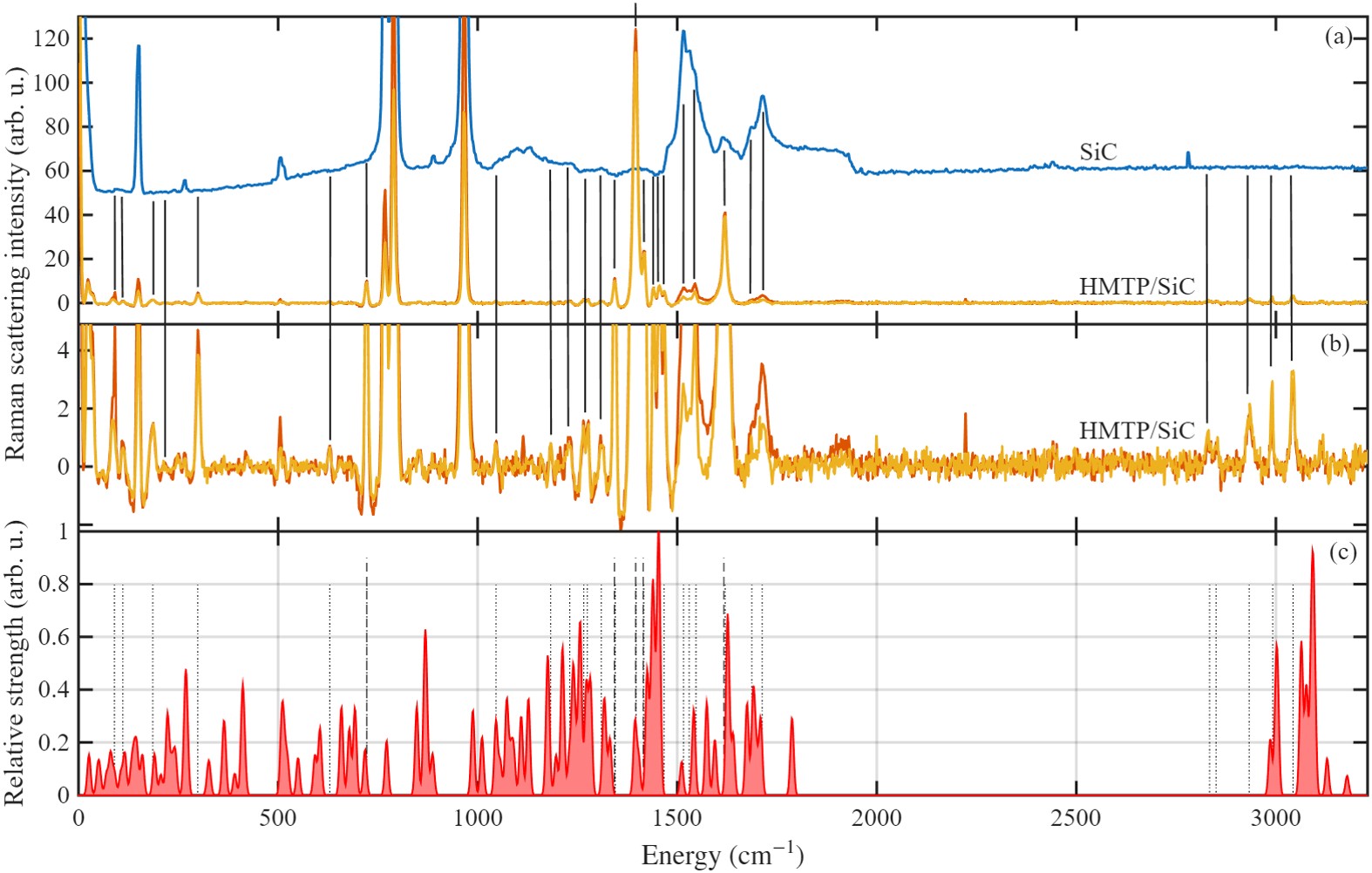}
\caption{Vibrational characterization and Molecular Dynamics (MD) validation. (a) Raman spectra of HMTP compared to the bare substrate ($\lambda_\text{exc}=532$~nm). Spectra from the SiC substrate (blue) and the HMTP film at two distinct spatial locations (red and orange) are shown. (b) Magnified view of the low-intensity region, highlighting the subtle molecular vibrational features. Vertical solid lines denote characteristic HMTP modes; peaks overlapping with the blue SiC reference represent substrate phonon modes. (c) Calculated Vibrational Density of States (VDOS) of HMTP obtained from MD simulations. Simulated transitions are convoluted with a Gaussian broadening function to match experimental resolution. Vertical dashed and dotted lines correlate the primary and secondary experimental modes with the simulated VDOS peaks.}
\label{FigRaman}
\end{figure*} 

\begin{table*}
\caption{\label{tab:Raman} Raman Peak Positions, Intensities, and Tentative Assignments for HMTP on SiC/graphene.}
\begin{ruledtabular}
\begin{tabular}{S[table-format=4.2] S[table-format=3.2] l c}
{Frequency (\si{\cm^{-1}})} & {Intensity (a.u.)} & {Suggested Assignment} & {Ref.} \\
\hline
88  & 1.7   & Lattice/Intermolecular Phonon & \cite{DellaValleJPCB2004,GirlandoJCP2011} \\
110 & 1.1   & Lattice/Intermolecular Phonon & \cite{DellaValleJPCB2004,GirlandoJCP2011} \\
185 & 1.4   & Whole-molecule Skeletal Bending & \cite{TroisiJPCA2006}  \\
297 & 4.4   & Methoxy -OCH$_3$ deformation & \cite{MeganathanJRS2010,BinoyJRS2004} \\
632 & 0.8   & In-plane Phenyl Ring Deformation & \cite{LiuBJ1990} \\
722 & 9.6   & In-plane Phenyl Ring Deformation, out-of-plane C-H deformation & \cite{LiuBJ1990,Bellamy1975} \\
1047 & 1.1  & C-O-C stretching & \cite{Socrates1980} \\
1183 & 0.5  & C-H in-plane bending & \cite{Socrates1980} \\
1232 & 1.1  & Ring-Methoxy stretching & \cite{Socrates1980} \\
1266 & 1.4  & Ring-Methoxy stretching & \cite{Socrates1980} \\
1274 & 1.4  & Ring-Methoxy stretching &  \cite{Socrates1980}\\
1310 & 1.1  & Softened Kekulé distorsion in $S_1$/C-O-C stretching & \cite{KunishigeJCP2013}  \\
1343 & 11 & Kekulé distorsion (FC Active, 168 meV) & \cite{KunishigeJCP2013} \\
1396 & 120 & C–C stretching & \cite{ZhangAJ2010} \\
1415 & 24 & Symmetric CH$_3$ Deformation & \cite{Socrates1980} \\
1438 & 5.7  & CH$_3$ Deformation & \cite{Socrates1980} \\
1457 & 7.8  & CH$_3$ Deformation & \cite{Socrates1980} \\
1470 & 5.7  & Asymmetric CH$_3$ Deformation &  \cite{Socrates1980} \\
1514 & 4.1  & Aromatic C=C Stretch & \cite{Socrates1980} \\
1529 & 6.0  & Aromatic C=C Stretch & \cite{Socrates1980} \\
1544 & 5.7  & Aromatic C=C Stretch & \cite{Socrates1980} \\
1617 & 42 & Primary Aromatic C=C Stretch & \cite{Socrates1980} \\
1685 & 1.7  & Combination Band / Overtone & -- \\
1714 & 3.5  & Combination Band / Overtone & -- \\
2832 & 0.8  & 2$\times$ Symmetric CH$_3$ Deformation &  \cite{Socrates1980} \\
2851 & 0.8  & Symmetric CH$_3$ Stretching &  \cite{Socrates1980}\\
2938 & 1.4  & 2$\times$ Asymmetric CH$_3$ Deformation &  \cite{Socrates1980} \\
2990 & 2.3  & Asymmetric CH$_3$ Stretching &  \cite{Socrates1980} \\
3045 & 3.5  & Aromatic C-H Stretching &   \cite{Socrates1980} \\
\end{tabular}
\end{ruledtabular}
\end{table*}

\begin{figure}[t!]
\centering
\includegraphics[width=8cm]{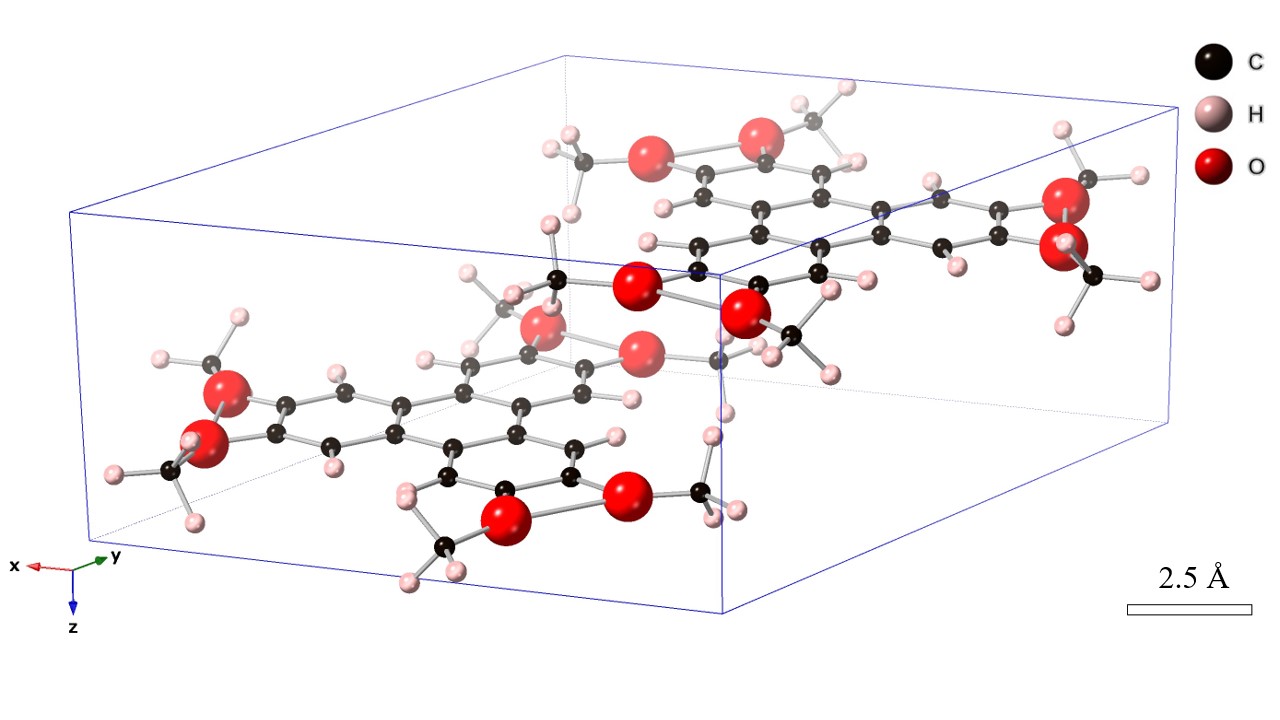}
\caption{Crystallographic unit cell of 2,3,6,7,10,11-hexamethoxytriphenylene (HMTP) is depicted by a blue rhombic prism. The spheres represent carbon (black), oxygen (red), and hydrogen (white) atoms. The arrangement corresponds to the $P6_3/m$ space group symmetry.}
\label{FigUnitCell}
\end{figure}
\subsection{Surface Purification and Residue Remediation}
While the optimized lithography minimizes initial contamination, the complete removal of ubiquitous resist residues is paramount for resolving intrinsic molecular signatures. Building upon the protocol proposed by Tyagi et al.~\cite{TyagiColettiNanoscale14-2167-2022}, we developed a refined wet-chemical cleaning procedure specifically designed to prevent the adsorption of polymer fragments. The conventional workflow typically involves distinct cleaning cycles separated by intermediate drying steps; however, we find that nitrogen drying before the final removal stage can lead to the permanent pinning of undissolved residues to the graphene lattice.

To circumvent this, we implement a continuous-wet protocol: the sample undergoes an initial immersion in acetone (2.5 h), followed by successive rinses in IPA (30 min) and deionized water (5 min), without any intervening drying stages. The sample is then transferred directly into a bath of AR 600-71 (a mixture of 1,3-dioxolane and 1-methoxy 2-propanol). In a significant departure from the 3-minute dwell time suggested in literature~\cite{TyagiColettiNanoscale14-2167-2022}, we found that an extended 1.5 h immersion in the dioxolane-based remover is necessary to fully solubilize the most resilient cross-linked PMMA/MMA remnants. Only after this exhaustive cleaning is the sample rinsed in DI water and dried under a compressed nitrogen flow. This modification ensures that the final surface—as validated by our subsequent LEEM and LEED analysis—retains the structural integrity of the epitaxial graphene while achieving a state of cleanliness typically reserved for in-situ Ultra-High Vacuum (UHV) preparation.

\subsection{Atomic Integrity and Molecular Templating}
\label{sec:AtomicIntergrity}
The structural integrity and atomic-scale cleanliness achieved via the never-dry dioxolane protocol were validated through UHV surface characterization of the active device regions. XRD studies confirm that the outgassing of the samples at elevated temperatures—required for LEEM/LEED characterization—does not modify the underlying graphene or metal contacts (see Appendix~\ref{sec:XRD} for XRD data before/after annealing and after molecular deposition). As shown in Fig.~\ref{FigLEED}(a), Low-Energy Electron Diffraction (LEED) displays sharp diffraction spots and a low diffuse background in the complete area of the graphene electrode, confirming an atomically clean surface successfully recovered from resist residues post-nanofabrication. The SiC(0001) diffraction spots are visible (orange arrows), alongside the six-fold symmetric ($1\times1$) spots characteristic of the epitaxial graphene lattice (yellow arrow) and the ($6\sqrt{3}\times6\sqrt{3}$)$\text{R}30^\circ$ surface reconstruction (violet arrows). This uniformity is corroborated by Low-Energy Electron Microscopy (LEEM) [Fig.~\ref{FigLEED}(d)], which reveals a high-contrast, homogeneous surface entirely devoid of the stochastic speckle contrast typical of cross-linked polymer remnants~\cite{PirkleAPL99-122108-2011}. This pristine, atomically flat landscape provides the necessary template for the subsequent self-assembly of the HMTP overlayer. Following submonolayer deposition, the LEED pattern [Fig.~\ref{FigLEED}(b)] reveals two additional sets of six-fold symmetric diffraction spots. These sets originate from two discrete, symmetry-equivalent alignments of the HMTP lattice with respect to the underlying graphene. The corresponding LEEM image [Fig.~\ref{FigLEED}(e)] shows the initial formation of HMTP aggregates (blue arrows). We note that the bright stripes (green arrows) originate from charging of the SiC substrate, where graphene was removed during oxygen plasma etching. To confirm the atomic structure, Scanning Tunneling Microscopy (STM) and its Fast Fourier Transform (FFT) are presented in Figs. \ref{FigLEED}(f) and (c), respectively. The FFT confirms a lattice constant of $a = 1.2(1)$ nm, consistent with the LEED diffraction pattern. The presence of only one hexagonal set in the FFT—compared to the two sets in LEED—is attributed to the localized probe area of STM, whereas LEED averages over larger domains containing both orientations. Building upon this well-defined interface, we grew 50 nm HMTP films on pre-patterned devices [Appendix~\ref{sec:Growth}]. The growth of 50~nm thick HMTP follows the first layer resulting in a highly ordered epitaxial film~\cite{DevanshuArxiv}. As illustrated by the optical micrograph [Fig.~\ref{FigDevice}(a)] and SEM [Fig.~\ref{FigDevice}(b)] and detailed AFM [Appendix~\ref{sec:Growth}, Fig.~\ref{FigAFMdetailSI}], the growth kinetics differ markedly between graphene, gold interdigital contacts, and the inter-contact area. AFM height images [Fig.~\ref{FigDevice}(c,d)] further reveal distinct morphological regimes; on the graphene areas, the HMTP maintains a higher degree of structural coherence, whereas growth on gold and SiC appears more disordered. This substrate-dependent morphology underscores the critical role of the graphene template in enforcing the long-range order necessary for intrinsic vibronic coupling.

\subsection{Vibrational Modes and Molecular Dynamics}
The vibrational landscape of the HMTP overlayer was characterized using a specialized Raman acquisition protocol designed to mitigate the extreme photosensitivity of the molecular species~\cite{AndersonAPB120-2015-Ramandifficulty,Gottfried2013}. To isolate the intrinsic molecular signatures from the dominant carbonaceous features of the substrate, measurements were performed on bare SiC regions [Fig.~\ref{FigRaman}(a)]; this avoids spectral overlap with the graphene D and G peaks, which otherwise obscure the characteristic HMTP lattice modes. To prevent photodegradation~\cite{SurovtsevaJAS2010}, we restricted the excitation power to the sub-milliwatt regime ($< 1$ mW). While individual spectra at this power level are dominated by shot noise, we recovered the full vibrational fingerprint by implementing a dynamic $x$-$y$ mapping protocol. By rapidly scanning a $30 \times 30$ $\mu\text{m}^2$ area and spatially averaging the resulting 900 spectra, we achieved the high signal-to-noise ratio necessary to resolve low-intensity Raman modes. The high reproducibility between traces from distinct sample areas [red and orange curves, Fig.~\ref{FigRaman}~(a,b)] confirms the exceptional chemical and structural uniformity of the molecular aggregates. To support these experimental signatures, we calculated the vibrational density of states (VDOS) using molecular dynamics (MD) simulations~\cite{CrystalMaker2025} [Fig.~\ref{FigRaman}(c)]. The simulation utilized a unit cell containing two inequivalent HMTP molecular positions, as identified in our structural analysis (Fig.~\ref{FigUnitCell}), with periodic boundary conditions used to emmulate the long-range order of the molecular crystal. The calculated vibrational energies exhibit excellent agreement with experimental data—typically within 10–20\% for MD~\cite{PiaPRL2024precision}. A comprehensive summary of the identified Raman modes, their relative intensities, and tentative assignments is provided in Table \ref{tab:Raman}. 

The vibrational spectrum of HMTP is dominated by two strong features at $1396 \text{ cm}^{-1}$, and $1617 \text{ cm}^{-1}$, corresponding to the aromatic ring stretching (C-C, and C=C vibrations, respectively). The weaker mode at $1343 \text{ cm}^{-1}$, assigned to the Kekulé distortion~\cite{KunishigeJCP2013}, acts as the primary mediator for electron-phonon coupling. Finally, the even weaker feature at $1310 \text{ cm}^{-1}$ is attributed to the softening of this Kekulé mode in the $S_1$ excited state, and it might be also contributed by the C–O–C stretching of the methoxy groups~\cite{Socrates1980}.

This vibrational characterization serves as the foundation for extracting the Huang-Rhys factors and parameterizing the exciton-phonon coupling within the Holstein Hamiltonian.
\begin{figure*}[t]
\centering
\includegraphics[width=16cm]{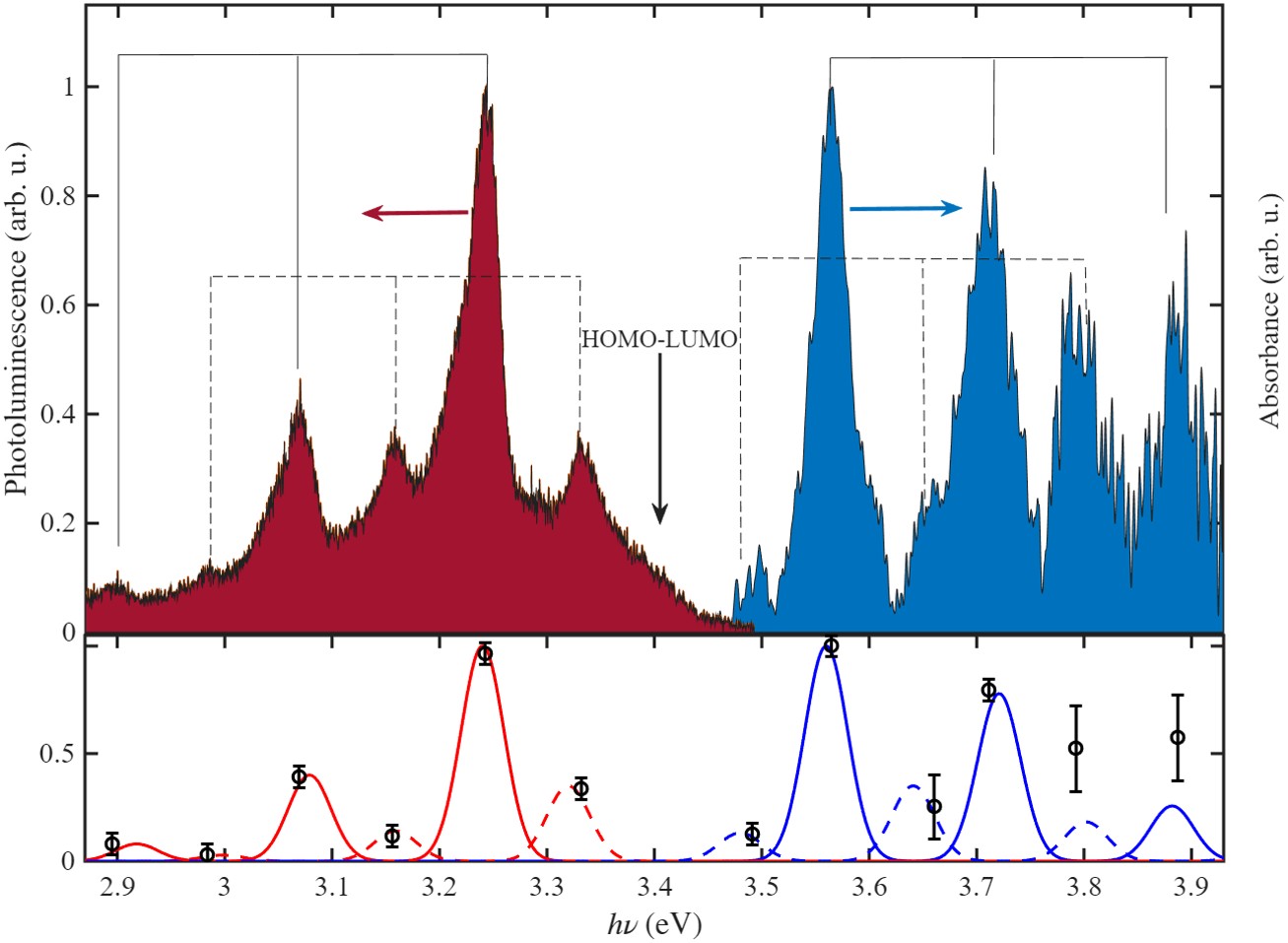}
\caption{(Top) Experimental photoluminescence (red shaded area) and absorption (blue shaded area) spectra. Solid vertical lines indicate the vibrational manifold of the primary Davydov branch, while dashed vertical lines denote the minor (secondary) branch. (Bottom) Theoretical simulation based on a Franck-Condon model incorporating Herzberg-Teller corrections. Solid and dashed curves represent the calculated transitions for the main and minor manifolds, respectively. The model is benchmarked against experimentally extracted peak positions and intensities (black open circles with error bars). The unperturbed monomer HOMO-LUMO transition ($E_{mon} = 3.4$ eV) is indicated for reference.}
\label{FigAbsEm}
\end{figure*} 

\begin{table*}[t!]
\centering
\caption{Experimental PL and Absorption Peak Positions and Assignments (Monomer ZPL = 3.40 eV)}
\label{tab:pl_abs_peaks}
\begin{ruledtabular}
\begin{tabular}{l S[table-format=1.3] S[table-format=1.3] l}
{Transition Type} & {Energy (eV)} & {Intensity (a.u.)} & {Assignment} \\
\hline
\textbf{Photoluminescence} & & & \\
Main Set (1-0) & 3.24 & 0.95 & Lower Davydov Branch ($S_1$) \\
Main Set (2-0) & 3.07 & 0.39 & 1st Phonon Replica \\
Main Set (3-0) & 2.90 & 0.08 & 2nd Phonon Replica \\
\addlinespace
\hline
Minor Set (1-0) & 3.33 & 0.33 & Upper Davydov Branch ($S_2$) \\
Minor Set (2-0) & 3.16 & 0.12 & 1st Phonon Replica \\
Minor Set (3-0) & 2.98 & 0.03 & 2nd Phonon Replica \\
\midrule
\textbf{Absorption} & & & \\
\hline
Main Set (1-0) & 3.57 & 0.99 & Upper Davydov Branch ($S_2$) \\
Main Set (2-0) & 3.71 & 0.78 & 1st Phonon Replica \\
Main Set (3-0) & 3.89 & 0.57 & 2nd Phonon Replica \\
\addlinespace
\hline
Minor Set (1-0) & 3.49 & 0.12 & Lower Davydov Branch ($S_1$) \\
Minor Set (2-0) & 3.66 & 0.25 & 1st Phonon Replica \\
Minor Set (3-0) & 3.79 & 0.52 & 2nd Phonon Replica \\
\bottomrule
\end{tabular}
\end{ruledtabular}
\end{table*}

\subsection{Electronic Structure and Excitonic Manifold}
\label{sec:ElectronicStructure}
\begin{figure}[t!]
\centering
\includegraphics[width=8.5cm]{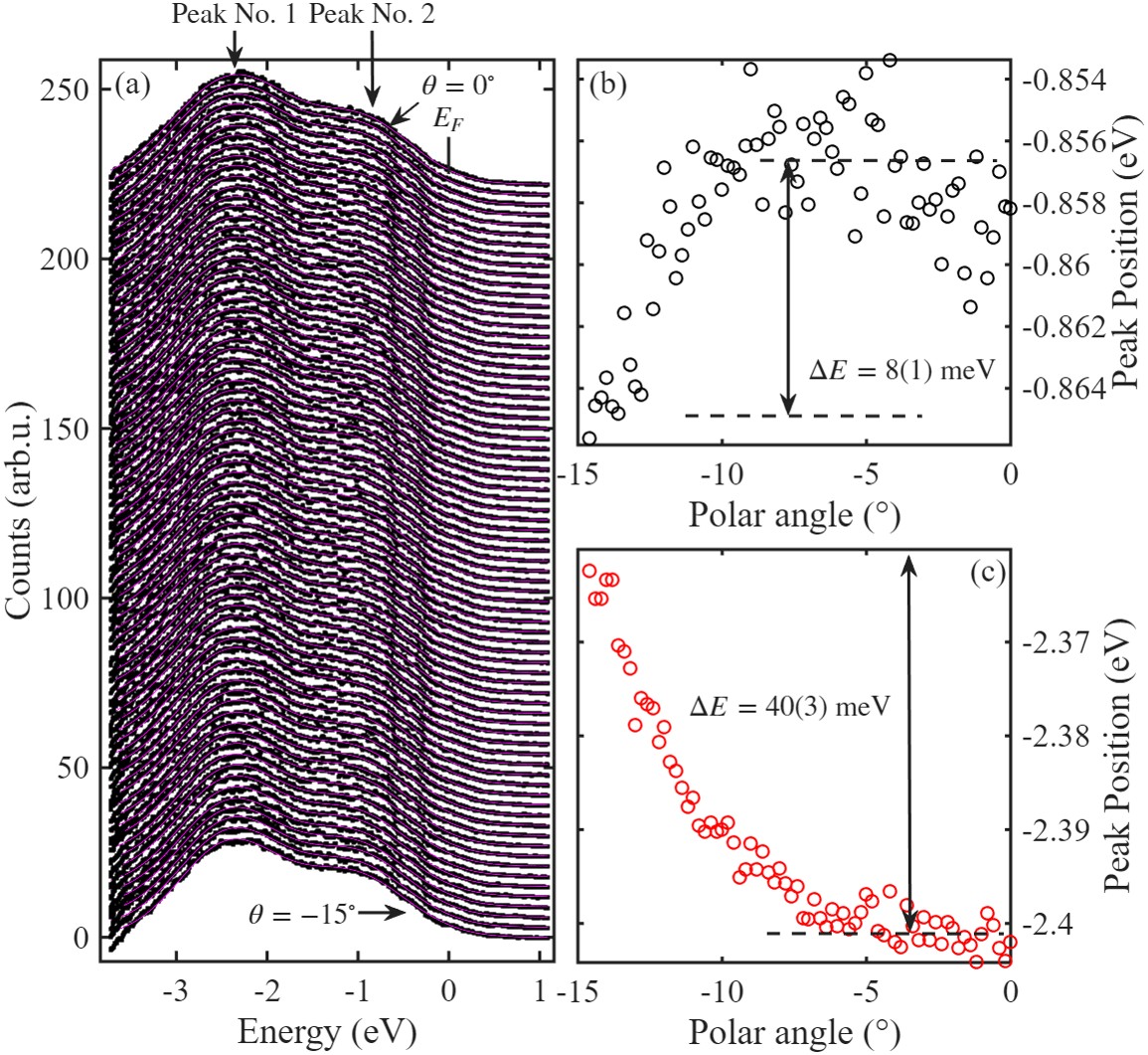}
\caption{Electronic band structure and binding energy analysis of HMTP on graphene. (a) ARPES intensity map acquired along the $\Gamma-K$ high-symmetry direction of the underlying graphene lattice; the energy axis is referenced to the Fermi level ($E - E_F = 0$ eV). (b, c) Detailed lineshape analysis of the two primary photoemission features located at (b) $E - E_F = -0.86$ eV and (c) $E - E_F = -2.40$ eV. Discrete peak positions were extracted using a center-of-mass integration method to track the dispersion (or lack thereof) across the sampled $k$-space.}
\label{FigARPES}
\end{figure}

The excitonic landscape of the HMTP-graphene interface was probed using room-temperature photoluminescence (PL) and absorption spectroscopy (Fig.~\ref{FigAbsEm}). PL was excited at $\lambda_\text{exc}=355$~nm using low excitation power densities to ensure the absence of photo-induced degradation. Absorption profiles were extracted via Fourier Transform Photocurrent Spectroscopy using the underlying SiC substrate as a highly sensitive internal probe. By normalizing the photo-current response of the functionalized device against the pristine SiC reference, we resolved the molecular absorption features. To account for minor setup variations during sample transfer, a smoothly varying spline background was subtracted; we verified that this procedure preserved the integrity of the spectral line shapes without introducing artificial optical responses. Both the PL and the absorption spectra exhibit a striking mirror symmetry around the HMTP monomer Highest Occupied Molecular Orbital-to-Lowest Unoccupied Molecular Orbital (HOMO-LUMO) transition, characterized by two distinct sets of transitions: a primary, high-intensity manifold and a secondary, weaker set of satellite peaks (Table \ref{tab:pl_abs_peaks}). To contextualize these optical signatures within the broader electronic structure, we performed ARPES on the HMTP overlayer (Fig.~\ref{FigARPES}). Analysis of the two major valence levels was conducted using both multi-peak Gaussian fitting and a model-unbiased Center-of-Mass (COM) integration to track the energy centroids [Fig.~\ref{FigARPES}(c,d)]. The resulting dispersion is remarkably flat, with a maximum bandwidth ($BW$) of only $\approx 40$ meV. Within a tight-binding framework (see Appendix~\ref{sec:TB}), where $BW = 13t$ for the HMTP crystal geometry, we estimate the intermolecular electronic coupling ($t$) to be on the order of $1$--$5$ meV. This small electronic bandwidth—being significantly lower than the characteristic vibrational energies ($10$--$200$ meV) identified via Raman spectroscopy—firmly places the HMTP-graphene system in the strong electron-phonon coupling regime. In this limit, the traditional Bloch-wave description is renormalized into a set of discrete vibronic levels. The observed spectral features thus represent the transition from a non-interacting monomeric picture to the formation of Davydov-split branches, where intermolecular exchange interactions lift the degeneracy of the excitonic states into bright (bonding) and dark (antibonding) branches. 

\begin{figure*}[t!]
\centering
\includegraphics[width=14cm]{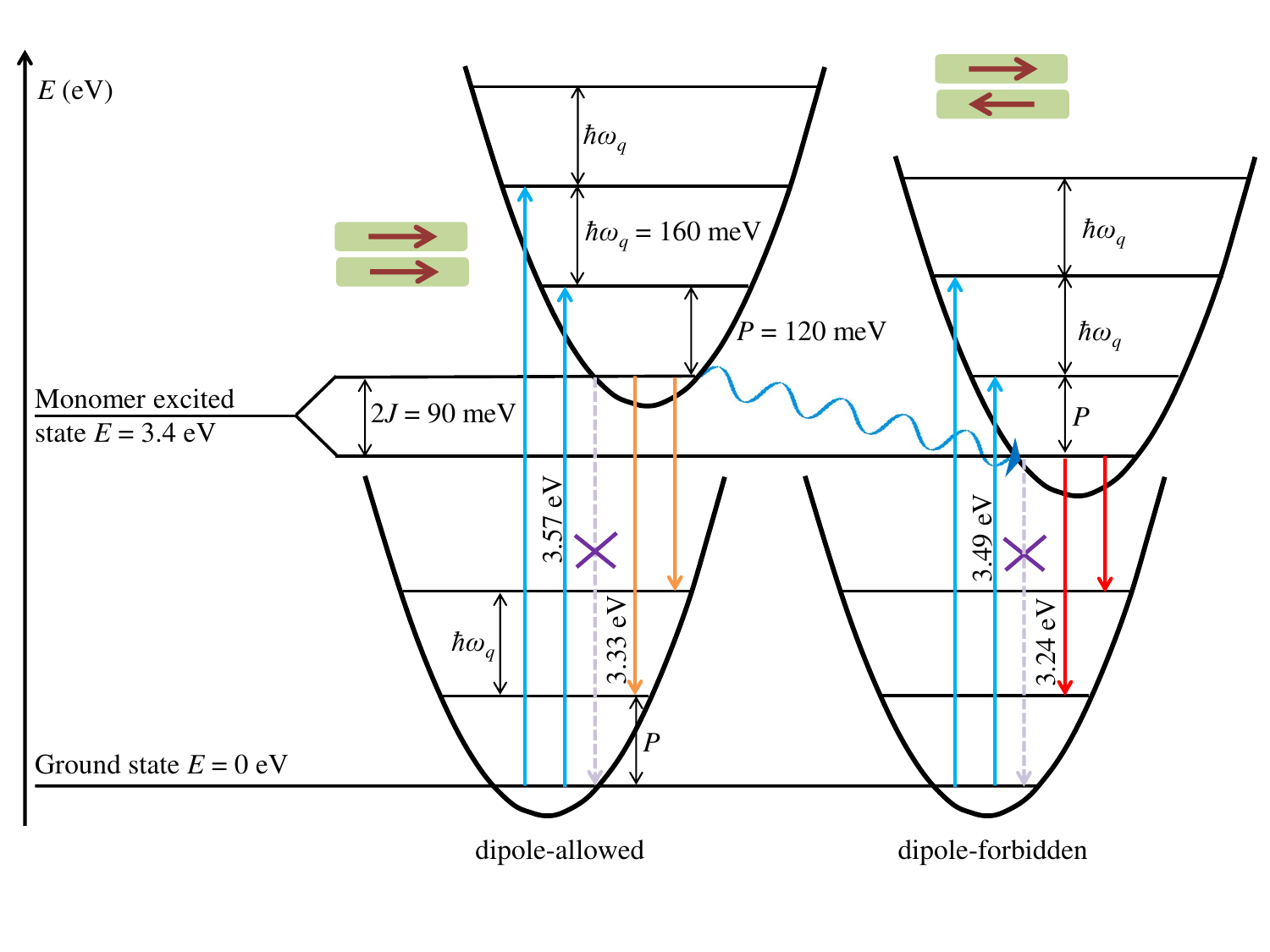}
\caption{Configurational coordinate diagram of the HMTP electronic states. Potential energy parabolas for the ground state ($S_0$) and the Davydov-split excited states ($S_1$). Vertical blue and orange/red arrows denote absorption and emission transitions, respectively. The relaxation pathway from the upper (bright) to the lower (dark/polaron) branch is indicated by the wavy line. The forbidden transitions 0-0 are depicted by the violet dashed lines (crossed).}
\label{FigEnergySchema}
\end{figure*} 
\subsection{Davydov Splitting}
The symmetry of the absorption and emission profiles relative to the monomer transition suggests a dominant Franck-Condon (FC) mechanism, modulated by the unique packing of the HMTP lattice. Given the intermolecular electronic coupling ($t \approx 5$ meV) is significantly smaller than the polarization energy ($P~\approx120$ meV), we describe the system using a framework of quasi-discrete vibronic levels within the strong-coupling regime~\cite{Ref3WestJPCB5157-2010}. As identified in the structural analysis (Fig.~\ref{FigUnitCell}), the HMTP unit cell contains two inequivalent molecular sites. This arrangement induces the formation of bonding and anti-bonding excitonic states, lifting the degeneracy of the HOMO-LUMO transition into two Davydov-split branches separated by $2J=90$~meV (Fig.~\ref{FigEnergySchema})~\cite{BeljonneJCP112-2000}. Unlike the randomized azimuthal distributions seen on amorphous supports~\cite{WittePSSA2008}, the graphene template enforces a macroscopic dipole alignment that allows us to resolve the redistribution of oscillator strength between these branches.

\subsection{Absorption and the Herzberg-Teller Correction}
Primary absorption, centered at 3.57 eV, corresponds to the dipole-allowed transition to the upper Davydov branch (1–0 transition). Higher-energy peaks at 3.71 eV and 3.89 eV represent subsequent phonon replicas (2–0, 3–0), characterized by the emission of $\approx 160(20)$ meV phonons—identified in our Raman data as the Kekulé distorsion of HMTP aromatic rings. A secondary, lower-intensity set of absorption peaks (dashed lines, Fig.~\ref{FigAbsEm}) is attributed to transitions to the lower Davydov branch. While these transitions are nominally symmetry-forbidden, the strong polaron coupling distorts the lattice sufficiently to transfer oscillator strength into these dark states. We find that the intensities of the main manifold follow a standard FC progression with a relatively small Herzberg-Teller (HT) correction $\alpha = 0.2(1)$ - see the Appendix~\ref{sec:FCaHTderivation} for derivation. However, the minor manifold requires a larger HT correction $\alpha = 0.8(3)$ to describe the anomalous intensity distribution.
\begin{equation}
I_n(S,\alpha) \propto \frac{S^n}{n!}\left(1 + \alpha\frac{n\pm S}{\sqrt{S}}\right)^2,  
\label{eq:IntensityFTaHTmaintext}
\end{equation}

where $S \approx 0.4$ is the Huang-Rhys factor and $\alpha$ accounts for the coordinate-dependence of the transition dipole moment and $+(-)$ stands for emission (absorption). This HT contribution highlights the non-Franck-Condon effects introduced by the intense exciton-phonon coupling in the ordered overlayer~\cite{Ref6TavazziJPCC118-8588-2014}.
\subsection{Emission and Hot Photoluminescence}
Surprisingly, the emission spectra are well-described by a pure FC mechanism ($\alpha \approx 0$), indicating that the HT distortion is primarily an excited-state phenomenon. Following lattice relaxation to the $n=0$ vibronic state, polarons relax to the lower Davydov branch (blue wavy line, Fig.~\ref{FigEnergySchema}). The resulting strongest PL transition at 3.24 eV represents the 1–0 emission from this darker branch (as dictated by Kasia's rule~\cite{KasiaAnnuRevPhysChem7-403-1956}), followed by the corresponding 160(20) meV phonon replicas. We also resolve a minor set of peaks starting at 3.33 eV, which we attribute to hot photoluminescence from the upper Davydov branch.

\subsection{Quantitative Spectral Modeling}

\begin{table*}[t!]
\centering
\caption{Parameters for the Vibronic Model of HMTP on SiC/Graphene}
\label{tab:vibronic_parameters}
\begin{ruledtabular}
\begin{tabular}{l l S[table-format=1.3] l}
\textbf{Category} & \textbf{Parameter} & \textbf{Value} & \textbf{Units/Notes} \\
\hline
Electronic Origin & Monomer Energy ($E_{mon}$) & 3.4 & \si{\eV}  \\
Crystal Shifts    & Polarization Shift ($P$) & \pm 120 & \si{\meV} \\
                  & Excitonic Coupling ($J$) & \pm 45 & \si{\meV} \\
\addlinespace
Vibronic Scaling  & Phonon Energy ($\hbar\omega$) & 160(20) & \si{\meV}  \\
                  & Phonon Quanta ($n$) & {0, 1, 2} & Integer  \\
                  & Huang-Rhys Factor ($S$) & 0.4(2) & Dimensionless \\
                  & Line Broadening ($\sigma$) & 20 & \si{\meV} (Gaussian) \\
\addlinespace
Branch Indices    & Branch Index ($m$) & {$\pm$ 1} & $+1$: Bright, $-1$: Dark \\
                  & Weighting Factor ($C_m$) & {Var.} & Relative branch population \\
\addlinespace
HT Correction     & Main PL Branch ($\alpha_\text{PL, main}$) & 0 & Vibronic-Induced Transition \\
($\alpha$)        & Minor PL Branch ($\alpha_\text{PL, minor}$) & 0 & Allowed (Hot) Transition \\
                  & Main Abs Branch ($\alpha_\text{Abs, main}$) & 0.2(1) & Weak Vibronic Mixing \\
                  & Minor Abs Branch ($\alpha_\text{Abs, minor}$) & 0.8(3) & Forbidden (Vibronic Induced) \\
\bottomrule
\end{tabular}
\end{ruledtabular}
\end{table*}

To illustrate the proposed vibronic framework, we modeled the PL and absorption spectra using a series of Gaussian functions with a fixed broadening of $\sigma = 20$ meV to account for the experimental linewidths. The total spectral response $A(E)$ is expressed as a sum over the two Davydov branches ($m = \pm$) and their associated phonon manifolds:
\begin{equation}
    A(E) = \sum_{m = \pm} \sum_{n = 0}^{\infty} \mathcal{C}_m \cdot I_n(S, \alpha) \cdot \exp\left( -\frac{(E - E_{n,m})^2}{2\sigma^2} \right),
\end{equation}
where the transition energies $E_{n,m}$ are governed by the interplay of the monomer HOMO-LUMO gap, the polarization energy $P$, and the intermolecular exchange coupling $J$ for absorption:
\begin{equation}
    E_{n,m} = E_{\text{monomer}} + P \pm J + n\hbar\omega,
\end{equation}
and emission
\begin{equation}
    E_{n,m} = E_{\text{monomer}} - P \pm J - n\hbar\omega.
\end{equation}
In this model, $P$ represents the energy required to reconfigure the lattice upon exciton formation (the polaron shift), while $J$ signifies the Davydov splitting between the bonding and anti-bonding branches. As shown by the red (PL) and blue (absorption) curves in the bottom part of Fig.~\ref{FigAbsEm}, this analytical approach provides an exceptional fit to the experimental data. The solid lines denote the primary dipole-allowed transitions, while the dashed lines represent the minor transitions whose intensities in absorption are dictated by the Herzberg-Teller corrected Huang-Rhys factors. The success of this fit confirms that the complex optical response of the HMTP-graphene interface can be rigorously described by a Holstein Hamiltonian parameterized solely by our independent Raman, ARPES, and structural measurements. 

Our ability to resolve these intrinsic relaxation pathways~\cite{HernandoPRL97-216403-2006} is a direct consequence of the interfacial order. While previous studies by Riss~\cite{RissCrommieACSNano8-5395-2014} and Chen~\cite{CrommieACSNano7-6123-2013} observed vibronic signatures on UHV-prepared graphene, our platform achieves a comparable level of atomic-scale definition, moreover, on a nanofabricated device. By demonstrating that the dioxolane never-dry protocol restores UHV-level LEED/LEEM purity after lithography, we bridge the gap between high-complexity device architectures and pristine surface-science models. This decoupling from the substrate minimizes lifetime broadening~\cite{RissCrommieACSNano8-5395-2014}, transforming the HMTP-graphene interface into a scalable, solid-state emulator for the Holstein Hamiltonian.

\section{Conclusion}
In summary, we have demonstrated an atomically well-defined hybrid platform that bridges the gap between scalable nanofabrication and pristine surface-science models. By implementing an all-methacrylate resist stack and a never-dry dioxolane purification protocol, we successfully recovered a UHV-equivalent graphene template on a functionalized device architecture. This structural precision allowed for the first experimental parameterization of the Holstein Hamiltonian in an HMTP-graphene system, resolved through the interplay of LEED, STM, and high-fidelity Raman mapping. Identification of the Davydov-split manifold—characterized by a splitting of $2J$ and modulated by strong exciton-phonon coupling—reveals a hierarchy of bright and dark excitonic states. Crucially, the presence of symmetry-forbidden transitions from the lower Davydov branch signifies the formation of heavy, localized polarons with significantly suppressed radiative decay rates. These dark excitons represent an ideal candidate for solid-state quantum storage, as their decoupling from the radiative continuum offers a pathway toward extended coherence lifetimes in molecular-scale devices. Our platform provides a versatile tool for the next generation of organic-inorganic quantum emulators. Future work will leverage the gating capabilities of our interdigitated architecture to investigate the dynamic tuning of the Davydov splitting through external electric fields. Furthermore, the ability to maintain such high interfacial purity on pre-patterned devices opens the door for incorporating these molecular emulators into integrated photonic circuits, potentially enabling on-chip, exciton-mediated quantum information processing at room temperature.

\begin{acknowledgements}
The financial support from the Czech Science Foundation (Grantová agentura České republiky) under project 22-04551S is gratefully acknowledged. CzechNanoLab Research Infrastructure (ID 90251), funded by MEYS CR, is gratefully acknowledged for the financial support of the measurements/sample fabrication. We also acknowledge fruitful discussions with P. Lipavský. D.V. and J.N. additionally acknowledge the support by the project Quantum Materials for Applications in Sustainable Technologies, Grant No. CZ.02.01.01/00/22\_008/0004572.
\end{acknowledgements}

\section{Data availability statement}
The data that support the findings of this article are openly available.

\appendix

\section{Detailed Nanofabrication Parameters}
\label{sec:Fab}
\begin{figure*}[t!]
\centering
\includegraphics[width=18cm]{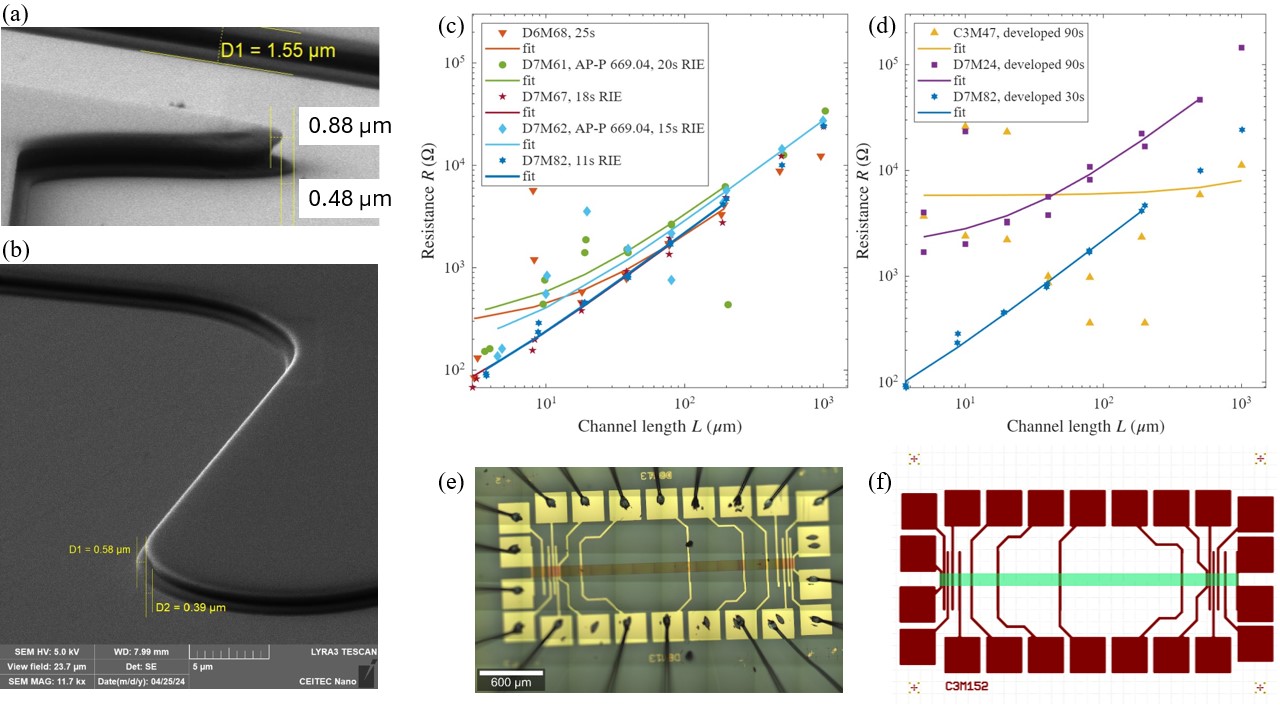}
\caption{Two-step e-beam lithography for clean, ohmic graphene edge contacts. (a) High-resolution SEM side-view ($85^\circ$ tilt) of the developed four-layer resist stack (from the SiC/graphene to the top layer - PMMA/MMA/MMA/Elektra), demonstrating the undercut profile required for metal lift-off. (b) Side-view SEM view of the developed resist over a large area. (c, d) Resistance vs. graphene channel length (TLM) for process optimization: (c) variation with oxygen plasma etching duration ($11\text{--}25$ s) and (d) comparison of resist development times ($60$ s vs. $90$ s). (e) Optical micrograph of the wire-bonded Transfer Length Method (TLM) device. (f) GDS layout illustrating the two-step process for the TLM device fabrication: (red) first-step metallization and (green) graphene shaping via oxygen plasma etching.}
\label{FigEdgeContactFab}
\end{figure*} 

\begin{table*}
\centering
\caption{Step-by-step spin-coating parameters for the four-layer resist stack.}
\label{tab:fabrication}
\begin{tabular}{lccc}
\hline
\hline
\textbf{Resist Layer} & \textbf{Spin Speed} & \textbf{Duration} & \textbf{Baking Temp. / Time} \\ \hline
PMMA (AR-P 679.04)  & 4000 rpm & 65 s & 200$^{\circ}$C / 10 min \\
MMA (AR-P 617.08) & 4000 rpm & 65 s & 200$^{\circ}$C / 10 min \\
MMA (AR-P 617.08) & 4000 rpm & 65 s & 180$^{\circ}$C / 10 min \\
Electra AR-PC 5090.02 & 4000 rpm & 65 s & 85$^{\circ}$C / 3 min \\ \hline
\hline
\end{tabular}
\end{table*}

\begin{table*}
\centering
\caption{Development and Cleaning Protocol for the All-Methacrylate Stack.}
\label{tab:development_cleaning}
\begin{tabular}{lll}
\hline
\hline
\textbf{Process Step} & \textbf{Chemical/Medium} & \textbf{Duration / Conditions} \\ \hline
\textit{Development:} & & \\
Electra Removal & DI Water & 30 s \\
Resist Development & AR 600-56 (Methyl isobutyl ketone) & 30 s \\
Stopper & Isopropyl Alcohol (IPA) & 15 s \\
Rinse & DI Water & 15 s \\
Drying & Compressed $N_2$ & Until dry \\ \hline
\textit{Post-Fab Cleaning:} & & \\
Primary Soak & Acetone & 2.5 h (Never-dry) \\
Rinse 1 & IPA & 30 min (Never-dry) \\
Rinse 2 & DI Water & 5 min (Never-dry) \\
Residue Removal & AR 600-71 (1,3-Dioxolane based) & 1.5 h (Never-dry) \\
Final Rinse & DI Water & 10 min (Never-dry)\\
Final Drying & Compressed $N_2$ & High-purity flow \\ \hline
\hline
\end{tabular}
\end{table*}
\subsection{Resist Stack Engineering}
The four-layer resist stack, as discussed in the main text~\ref{sec:Fabrication}, was engineered to create a specific sensitivity gradient to optimize the undercut profile for edge-contact formation while protecting the graphene channel (Table \ref{tab:fabrication}).

Sensitivity Gradient: The MMA (AR-P 617.08) layers are baked at different temperatures to modulate their sensitivity to the electron beam. The MMA layer baked at 200$^{\circ}$C becomes more sensitive than the upper MMA layer baked at 180$^{\circ}$C. This creates a controlled lateral development rate, resulting in a stable, high-fidelity undercut necessary for clean metal lift-off.

The protective PMMA Base: A bottom layer of PMMA (AR-P 679.04) is critical for graphene protection. Unlike MMA, this layer acts as a sacrificial barrier that protects the graphene from over-etching during the oxygen plasma step. Because oxygen plasma etching is not perfectly directional, the absence of this layer would allow the plasma to spread beneath the MMA undercut, over-etching the graphene beyond the intended contact area restricted by a more directional metal evaporation.

\subsection{E-beam Lithography and Optimization} Patterning was performed using a Raith 150-Two / Tescan Mira system at 20 kV with a beam current of 2–3 nA and a dose of 300 $\mu$C/cm$^2$ (step/line spacing: 50 nm). We found that the transition from ohmic to non-ohmic behavior is highly sensitive to both development and etching times. As shown in Fig.~\ref{FigEdgeContactFab}~c, oxygen plasma etching longer than 25 s results in non-ohmic contacts due to excessive lateral etching. Our optimal window was found to be 11–15 s. While the Shetty procedure uses sequential developers, our single-step development in AR 600-56 must be limited to 30 s. Extending development to 60 or 90 s, Fig.~\ref{FigEdgeContactFab}~d , causes the PMMA base layer to thin excessively and recede too far beneath the MMA undercut, disrupting the precise alignment of the metal evaporation with the graphene edge.

\section{Structural Stability of Metallization upon Thermal Annealing}
\label{sec:XRD}
To recover the pristine graphene surface for LEED and LEEM characterization, the fabricated devices must undergo a UHV outgassing procedure at 500$^{\circ}$C for 5 minutes~\ref{sec:AtomicIntergrity}. While the epitaxial graphene remains inherently stable at these temperatures, it is necessary to verify the structural integrity of the wirebonding pads and interdigitated metal contacts (Ti/Cu/Au stack) for the reference sample, to ensure that the device geometry remains intact for subsequent molecular deposition and electrical probing.
\begin{figure*}[t!]
\centering
\includegraphics[width=16cm]{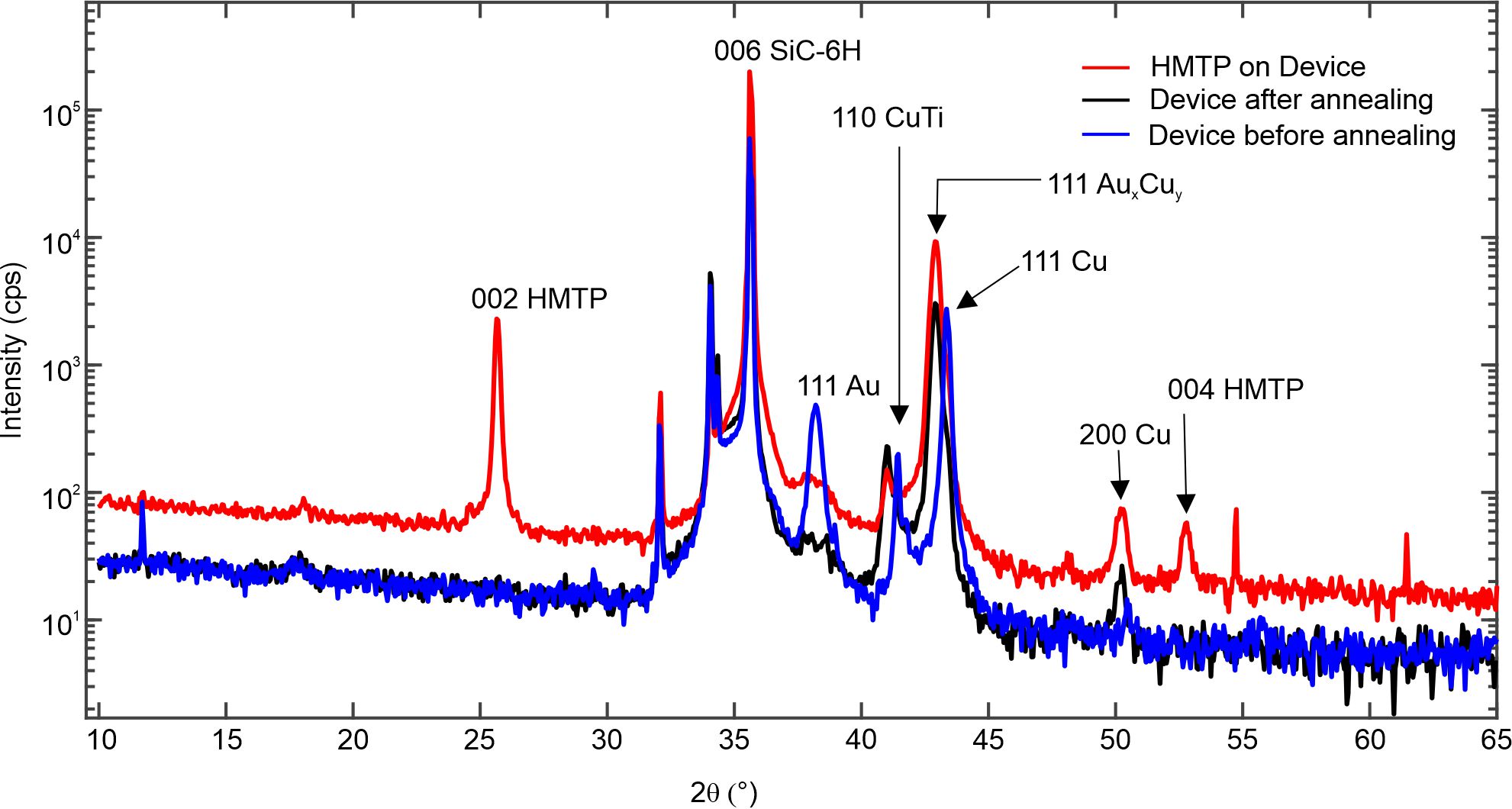}
\caption{Structural evolution of the Fourier Transform photocurrent device during processing. XRD diffractograms of the metallic interdigitated electrode device at different fabrication stages: initial device (black), following UHV outgassing at 500$^{\circ}$C for 5 min to remove adsorbed moisture (blue), and after 50 nm HMTP film deposition (red). Comparison of the black and blue curves reveals a broadening of the Au reflections and a slight shift/intensification of the Cu peaks, suggesting the onset of Au-Cu interfacial alloying during the thermal treatment. Despite these interfacial changes, the overall electrode geometry remains intact. The additional diffraction features in the red curve correspond to the crystalline phases of the HMTP layer.
}
\label{FigXRD}
\end{figure*} 
As shown in Figure \ref{FigXRD}, X-ray diffraction (XRD) was employed to monitor the evolution of the metal stack—comprising 5 nm Ti, 220 nm Cu, and 20 nm Au—throughout the processing stages. X-ray coplanar diffraction ($\theta-2\theta$) measurements were performed using a Rigaku SmartLab 3 X-ray diffractometer equipped with a rotating Cu anode (wavelength $\lambda=0.154$~nm) and a five-circle goniometer. For this measurement, a monochromatized and collimated incident X-ray beam with a size of 0.2~mm was achieved using a parabolic multilayer mirror, followed by a 0.3~mm pinhole and a 0.2~mm collimator. The diffracted intensity was recorded using a two-dimensional detector (HyPix-3000) positioned 150~mm from the sample.

In the as-deposited state, distinct Bragg reflections corresponding to the (111) Au and (200) Cu planes are observed. Following the 500$^{\circ}$C annealing step, the (111) Au peak exhibits a marked reduction in intensity and significant broadening, while the (200) Cu diffraction intensity increases. These spectral shifts are characteristic of inter-diffusional alloying at the Cu-Au interface, leading to the formation of a solid solution or disordered alloy phase.

Despite this interfacial alloying, the overall morphological integrity of the leads is preserved. This is evidenced by the persistent macroscopic definition of the contacts in optical micrographs and the successful nucleation of the molecular overlayer. Upon the deposition of HMTP on the annealed contacts, a new diffraction feature emerges at the position corresponding to the (002) reflection of the HMTP molecular crystal. This confirms that the annealed metal leads remain a suitable substrate for molecular assembly. We therefore conclude that while thermal processing induces a controlled Cu-Au alloying, it does not compromise the lithographic definition or the structural stability of the device contacts.

\begin{figure*}[t!]
\centering
\includegraphics[width=18cm]{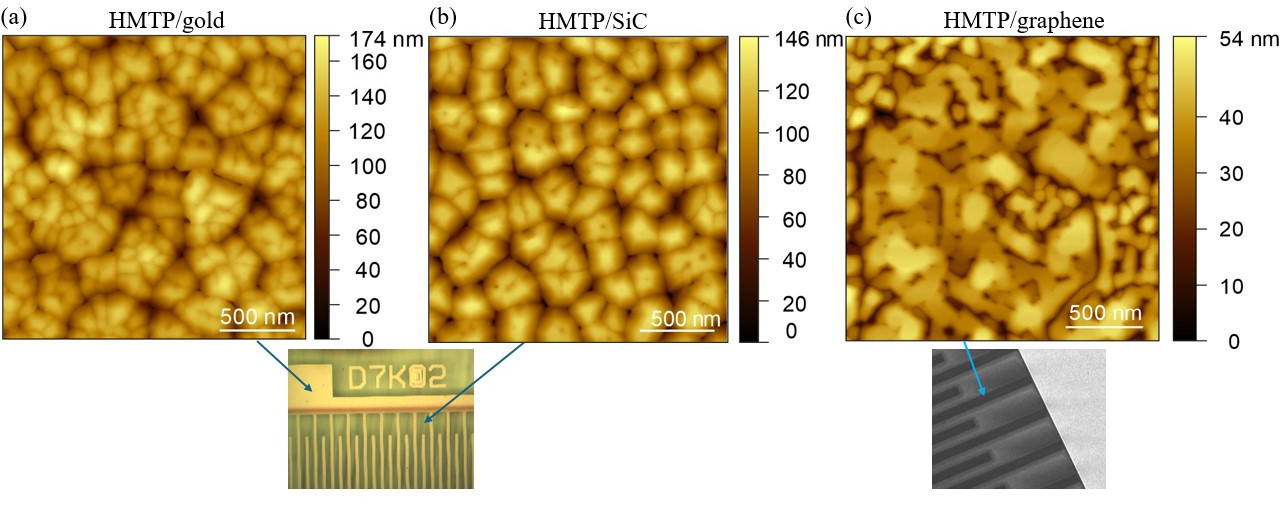}
\caption{Morphology of HMTP films on different device regions. AFM topography of a 50 nm thick HMTP film deposited on the Fourier Transform photocurrent device. (a) Surface morphology on the gold-covered electrode area ($20$ nm Au / $220$ nm Cu / $5$ nm Ti metallization). (b) Surface morphology in the channel region between the electrodes (SiC substrate area). Differences in grain size and nucleation density between (a) and (b) reflect the distinct surface energies of the polycrystalline gold contact versus the underlying substrate.}
\label{FigAFMdetailSI}
\end{figure*} 

\section{Graphene and HMTP Growth and Substrate-Dependent Morphology}   
\label{sec:Growth}
\subsection{Graphene Growth}
Epitaxial graphene was synthesized by thermal decomposition of the Si-face of 4H- or 6H-SiC(0001) substrates~\cite{BergerJPCB2004} (sourced from Coherent/II-VI Inc. and Wolfspeed/Cree Inc.). The substrates $5\times5$ mm$^2$ were cleaned in acetone, isopropanol, and deionized (DI) water prior to loading them into the growth furnace. The growth process followed a high-temperature annealing in a purified argon atmosphere (at a pressure of $\approx 1000$ mbar) to regulate the silicon sublimation rate~\cite{EmtsevNatMat2009,KuncPRA2017}. The thermal cycle consisted of sample outgassing and surface preparation at 800$^{\circ}$C and 1100$^{\circ}$C for 10 minutes each to ensure surface cleanliness and reconstruction. The final growth stage was taken at 1625-1650$^{\circ}$C for 5 minutes. The precise growth temperature is calibrated according to the aging of the graphite crucible, as established in our previous optimization studies~\cite{KuncPRA2017, ShestopalovJCG2025}. This calibration ensures the formation of a uniform monolayer with minimal bilayer coverage, providing the necessary atomically flat template for subsequent HMTP self-assembly.

\subsection{Submonolayer HMTP Growth}
All graphene samples (with and without contacts) were annealed at 530$^{\circ}$C for 20~min in a preparation chamber with a base pressure of (base pressure of 2$\times 10^{-10}$~mbar). For thin film experiments, HMTP molecules were deposited in the LEEM chamber using an organic material evaporator (MBE Komponenten, Quad Cell OEZ40). The molecular powder (Merck) was kept in a resistively heated quartz crucible and operated at 170-190$^{\circ}$C. The LEEM chamber design results in a deposition rate that is an order of magnitude lower than in thick-layer deposition.
Room-temperature Scanning Tunneling Microscopy (STM) images were obtained on a commercial system, Aarhus SPECS, using the KolibriSensor and a base pressure in the chamber of 1$\times 10^{-10}$~mbar. Measurements were performed on a GrSiC sample without contacts using a constant current mode, the sample bias of 0.7~V, and the tunneling current of 50~pA. The contrast of the images was adjusted in Gwyddion software.
Specs FE-LEEM P90 instrument was used to perform Low-Energy Electron Microscopy/Diffraction (LEEM/LEED) experiments (base pressure of 2$\times 10^{-10}$~mbar). The bright-field images were measured by collecting electrons from the central (0,0) beam. The diffraction patterns were taken from a surface area of $15×10~ \mu m^2$ or using a set of apertures that reduced the beam spot on the surface to 185~nm in diameter.

\subsection{Thin HMTP layers}
HMTP molecules were purchased from Merck. HMTP thin films with a thickness of 50 nm were deposited using an organic molecular beam evaporator (MBE Komponenten OEZ) equipped with a resistively heated quartz crucible operated at 175$^{\circ}$C. The deposition was carried out in a deposition chamber with a base pressure of $5\times10^{-10}$~mbar. Prior to deposition, the substrates were cleaned by annealing at 480$^{\circ}$C for 5-15 minutes under ultra-high vacuum (UHV) to remove surface contaminants.  The deposition rate, calibrated on pure single-layer graphene (SLG) samples using X-ray diffraction (XRD), was 5~$\mathrm{\AA}$/s. All the samples were kept at 25$^{\circ}$C (room temperature) during deposition, and identical growth conditions were applied for all thin film preparations.

To investigate the influence of the underlying substrate on the growth of 50~nm thin HMTP layer, discsussed in~\ref{sec:AtomicIntergrity}, we performed high-resolution AFM topography on $2\times2\ \mu\text{m}$ regions of the functional device (Fig.~\ref{FigAFMdetailSI}). This scale allows for a quantitative comparison of the grain boundary density and nucleation kinetics across the different material interfaces. AFM measurements were performed using a Bruker Dimension Icon microscope operated in tapping mode, and the acquired data were analyzed using Gwyddion software~\cite{Klapetek}. On the polycrystalline gold electrodes [Fig.~\ref{FigAFMdetailSI}(a)], the HMTP film forms a dense network of small, isotropic grains. Such a disordered landscape is expected to accelerate excitonic dephasing and localize transitions. In contrast, the morphology in the inter-electrode regions [Fig.~\ref{FigAFMdetailSI}(b)]—where the SiC is exposed—exhibits a distinct grain evolution. While more ordered than the growth on gold, it lacks the long-range azimuthal orientation observed on the pristine graphene template~\ref{FigAFMdetailSI}(c). This comparison confirms that the $1\times1$ graphene lattice is uniquely capable of suppressing stochastic nucleation and enforcing the structural coherence.

\section{Macroscopic Device Alignment and HMTP Film Distribution}
\label{sec:MakroDevice}
\begin{figure}[t!]
\centering
\includegraphics[width=7cm]{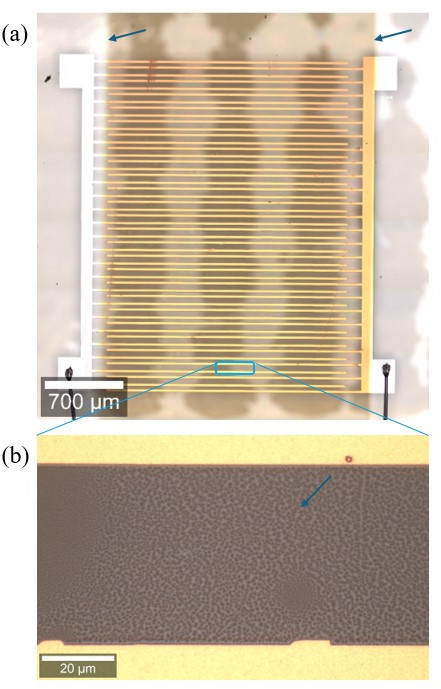}
\caption{Large-scale device morphology and HMTP film distribution. (a) Low-magnification optical micrograph of the complete Fourier Transform photocurrent device. The dark transverse region (edges indicated by blue arrows) defines the 50 nm thick HMTP layer deposited via thermal evaporation through a shadow mask. (b) High-magnification optical micrograph of the active channel area. The visible contrast (representative features highlighted by the blue arrow in (b) confirms the distribution of HMTP molecular aggregates between the interdigitated metal electrodes, demonstrating uniform coverage across the device geometry.}
\label{FigOpticalImageOfHMTP}
\end{figure} 
To confirm the spatial uniformity and precise alignment of the molecular overlayer with the underlying contact architecture, we performed large-area optical microscopy. The 50 nm thick HMTP film was deposited by thermal evaporation through a shadow mask, resulting in a well-defined molecular band across the active region of the device.

As shown in Figure \ref{FigOpticalImageOfHMTP}(a), the low-magnification micrograph reveals a clear contrast difference between the pristine substrate and the deposited HMTP band (shown by blue arrows). This macro-scale verification ensures that the interdigitated electrode array is fully integrated within the functional molecular region. At higher magnification [Fig.~\ref{FigOpticalImageOfHMTP}(b)], the distribution of HMTP molecular aggregates is visible across the active channel. The consistent optical contrast between the metal electrodes and the SiC regions confirms that the deposition protocol allows for continuous macroscopic coverage without compromising the lithographic definition of the device leads.

\section{Tight-Binding Derivation of the HMTP Band Structure}
\label{sec:TB}
\begin{figure*}[t!]
\centering
\includegraphics[width=15cm]{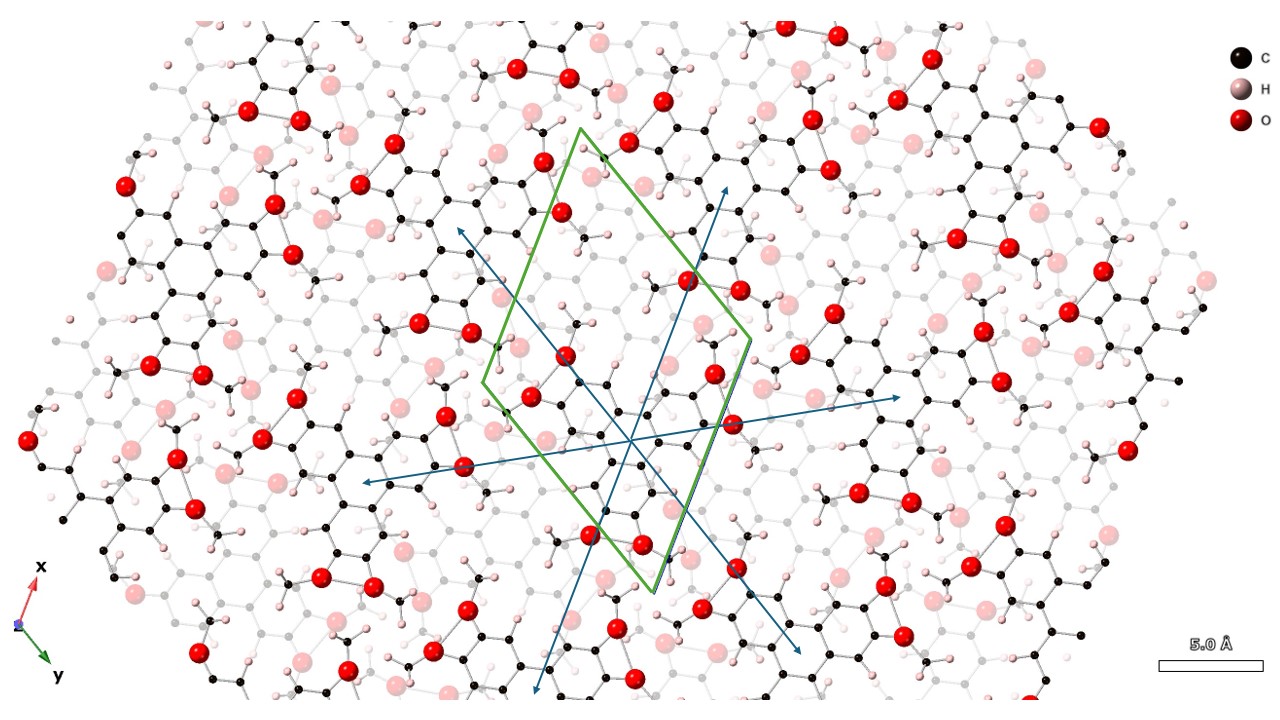}
\caption{In-plane molecular packing and nearest-neighbor coupling. Top-down view of the HMTP crystal lattice highlighting the intra-sublattice arrangement. Two molecular layers are shown, with the underlying layer (sublattice $\mathcal{B}$) shaded in grey to distinguish it from the primary layer (sublattice $\mathcal{A}$). The hexagonal unit cell is indicated by the green rhombus. The six nearest-neighbor vectors within the $ab$-plane are denoted by blue arrows, representing the primary hopping pathways for the intra-sublattice interaction ($t_{ab}$). These vectors correspond to the six terms in the intra-sublattice sum $\sum_{j} e^{i\mathbf{k}\cdot\mathbf{R}_j}$ used in the tight-binding Hamiltonian. Atomic species are indicated by black (C), red (O), and white (H) spheres.}
\label{FigProTB}
\end{figure*} 

To estimate the intermolecular coupling strength $t$ from our ARPES data, as we did in section~\ref{sec:ElectronicStructure}, we employ a nearest-neighbor tight-binding (TB) model based on the $P6_3/m$ crystalline symmetry of HMTP. The unit cell contains two inequivalent molecular sites, defining two sublattices ($\mathcal{A}$ and $\mathcal{B}$) as illustrated in Fig.~\ref{FigProTB}.
\subsection{Hopping Vectors and Coordination}
We consider the eight primary hopping pathways for an exciton or charge carrier: six intra-sublattice neighbors within the $ab$-plane and two inter-sublattice neighbors along the $c$-axis stacking direction. For intra-sublattice neighbors ($\mathcal{A} \rightarrow \mathcal{A}$ or $\mathcal{B} \rightarrow \mathcal{B}$), each HMTP molecule has six neighbors in the hexagonal plane at distance $a$:
    \begin{align}
        \mathbf{R}_{1,2} &= \pm a (1, 0, 0) \\
        \mathbf{R}_{3,4} &= \pm a \left( -\frac{1}{2}, \frac{\sqrt{3}}{2}, 0 \right) \\
        \mathbf{R}_{5,6} &= \pm a \left( -\frac{1}{2}, -\frac{\sqrt{3}}{2}, 0 \right)
    \end{align}

For inter-sublattice neighbors ($\mathcal{A} \rightarrow \mathcal{B}$), the two neighbors along the $c$-axis are at:
\begin{equation}
    \mathbf{R}_{7,8} = \pm \left( 0, 0, \frac{c}{2} \right).
\end{equation}

\subsection{Tight-Binding Hamiltonian}
The electronic states are described by the $2 \times 2$ Hamiltonian matrix in the basis of the two sublattices:
\begin{equation}
    \mathcal{H}(\mathbf{k}) = 
\begin{pmatrix}
H_{AA}(\mathbf{k}) & H_{AB}(\mathbf{k}) \\
H_{BA}(\mathbf{k}) & H_{BB}(\mathbf{k})
\end{pmatrix}
\end{equation}
The diagonal terms, representing the intra-sublattice dispersion, are obtained by summing the phase factors over the six in-plane neighbors:
\begin{equation}
    H_{AA}(\mathbf{k}) = \epsilon_0 - 2t_{ab} \left[ \cos(k_x a) + 2 \cos\left(\frac{k_x a}{2}\right) \cos\left(\frac{\sqrt{3} k_y a}{2}\right) \right]
\end{equation}
with
\begin{equation}
     H_{AA}(\mathbf{k}) = H_{BB}(\mathbf{k}).    
\end{equation}

The off-diagonal term represents the coupling between the two sublattices along the stacking direction:
\begin{equation}
   H_{AB}(\mathbf{k}) = H_{BA}^\ast(\mathbf{k}) = - 2t_{z} \cos\left(\frac{k_z c}{2}\right) 
\end{equation}

\subsection{Dispersion and Bandwidth}
Diagonalizing $\mathcal{H}(\mathbf{k})$ yields the energy eigenvalues
\begin{equation}
    E_{\pm}(\mathbf{k}) = H_{AA}(\mathbf{k}) \pm |H_{AB}(\mathbf{k})|.
\end{equation}

The total bandwidth ($BW$) is defined by the difference between the global maximum and minimum of the energy bands. For a simplified isotropic case where the hopping integrals are comparable ($t_{ab} \approx t_z = t$), the extrema occur at the $\Gamma$ point ($E_{min}$) and the $M$ or $K$ points ($E_{max}$). At the $\Gamma$ point ($\mathbf{k}=0$):$$H_{AA}(0) = \epsilon_0 - 6t, \quad H_{AB}(0) = -2t \implies E_{min} = \epsilon_0 - 8t.$$ At the zone boundary (namely at the $K$ point where $H_{AA} = \epsilon_0 + 3t$ and $k_z$ allows $H_{AB} = +2t$):
\begin{equation}
   E_{max} = \epsilon_0 + 5t 
\end{equation}

Thus, the total theoretical bandwidth is:
\begin{equation}
    BW = E_{max} - E_{min} = 13t
\end{equation}
This linear relationship allows us to directly map the $\approx 40$ meV dispersion observed in ARPES to an intermolecular coupling strength $t \approx 3$ meV, confirming the heavy-exciton, strong-coupling regime.


\section{Vibronic Transitions and Herzberg-Teller Corrections}
\label{sec:FCaHTderivation}
In this section, we provide a formal derivation of the vibronic intensity profiles, Eq.~(\ref{eq:IntensityFTaHTmaintext}), based on the Franck-Condon (FC) principle and the first-order Herzberg-Teller (HT) correction. We model the system using a displaced harmonic oscillator (DHO) manifold, where the ground and excited states are represented by linear harmonic oscillators (LHO) with a relative displacement in the generalized nuclear coordinate $Q$.

\subsection{Displaced Harmonic Oscillator}
The generalized coordinate $Q$ for a phonon mode with frequency $\omega$ and effective mass $M_{\text{eff}}$ is expressed in terms of the bosonic creation ($a^\dagger$) and annihilation ($a$) operators:
\begin{equation}Q = \sqrt{\frac{\hbar}{2M_{\text{eff}}\omega}}(a + a^\dagger).\label{eq:Q_op}\end{equation}
The displacement of the excited state potential energy surface relative to the ground state is given by $\Delta Q = \sqrt{2\hbar S / M_{\text{eff}}\omega}$,
where $S$ is the dimensionless Huang-Rhys factor. We utilize the displacement operator $D(\sqrt{S})$~\cite{FeynmanPhysRev.84.108} to map the ground state wavefunction onto the excited state manifold:
\begin{equation}D = e^{\sqrt{S}(a^\dagger - a)} = e^{-S/2} e^{\sqrt{S} a^\dagger} e^{-\sqrt{S} a},\end{equation}
where the second equality follows from the Baker-Campbell-Hausdorff identity. The operator $D$ acts on the operators as $D^\dagger a D = a + \sqrt{S}$ and $D^\dagger a^\dagger D = a^\dagger + \sqrt{S}$, shifting the expectation value of the position:
\begin{equation}\langle S | Q | S \rangle = \langle 0 | D^\dagger Q D | 0 \rangle = \sqrt{\frac{2\hbar S}{M_{\text{eff}}\omega}}.
\end{equation}

\subsection{Franck-Condon Intensities}
The transition amplitude in the Franck-Condon approximation assumes a constant electronic transition dipole moment $\mu_0$. The intensity $I_{0\to n}$ of the $n$-th vibronic replica is proportional to the square of the overlap integral (Franck-Condon factor):
\begin{equation}
D|0\rangle = e^{-S/2} \sum_{k=0}^\infty \frac{S^{k/2}}{\sqrt{k!}} |k\rangle.
\label{eq:Dna0}
\end{equation}
The overlap $\langle n | D | 0 \rangle$ yields the well-known Poissonian distribution for intensity:
\begin{equation}I_{0\to n} \propto |\mu_0 \langle n | D | 0 \rangle|^2 \propto e^{-S} \frac{S^n}{n!}.\label{eq:FC_final}
\end{equation}

\subsection{Herzberg-Teller Correction}
To account for the coordinate dependence of the transition dipole moment, we expand $\mu(Q)$ to the first order~\cite{KunduJPCB2022}:
\begin{equation}
    \mu(Q) = \mu_0 + \mu' Q,
\end{equation}

where $\mu' = (\partial \mu / \partial Q)_{Q_0}$. The total transition matrix element $M$ becomes:

\begin{equation}
    M = \mu_0 \langle n | D | 0 \rangle + \mu' \langle n | Q D | 0 \rangle.
\end{equation}

Using Eq.~(\ref{eq:Q_op}), the second term is evaluated as:

\begin{equation}
\langle n | Q D | 0 \rangle = \sqrt{\frac{\hbar}{2M_{\text{eff}}\omega}} \left( \langle n | D a^\dagger | 0 \rangle  + 2 \sqrt{S}\langle n | D | 0 \rangle   \right).
\label{eq:nQD0}
\end{equation}
Following the bosonic algebra for $\langle n | D | 1 \rangle$, we first identify how the displacement operator acts on the state $\ket{1}$.
\begin{equation}
\begin{split}
    D\ket{1}=e^{-\frac{S}{2}}e^{\sqrt{S} a^+}e^{-\sqrt{S} a}\ket{1} & = \\
= e^{-\frac{S}{2}}e^{\sqrt{S} a^+}\sum_{k=0}^\infty \frac{(-\sqrt{S}a)^k}{k!}\ket{1} & = \\ = e^{-\frac{S}{2}}e^{\sqrt{S} a^+} (\ket{1}-\sqrt{S}\ket{0})    \\
\end{split}
\end{equation}
We used the observation that in the sum; only the first two terms, $k=\{0,1\}$, are non-zero. We can apply the second part of the displacement operator, containing the creation operator
\begin{equation}
\begin{split}
    D\ket{1}=e^{-\frac{S}{2}}e^{\sqrt{S} a^+} (\ket{1}-\sqrt{S}\ket{0}) & =   \\
= e^{-\frac{S}{2}}\sum_{k=0}^\infty\frac{(\sqrt{S}a^+)^k}{k!} (\ket{1}-\sqrt{S}\ket{0}). \\  
\end{split}
\end{equation}
Using $(a^+)^k\ket{0}=\sqrt{k!}\ket{k}$, and $(a^+)^k\ket{1}=\sqrt{(k+1)!}\ket{k+1}$, we have
\begin{equation}
\begin{split}
     D\ket{1}= &  \\ e^{-\frac{S}{2}} & \sum_{k=0}^\infty\left [ \frac{S^{k/2}}{k!}\sqrt{(k+1)!}\ket{k+1}- \sqrt{S}     \frac{S^{k/2}}{k!}\sqrt{k!}\ket{k} \right].
\end{split}
\end{equation}
Noticing that the second term equals $-\sqrt{S}D\ket{0}$, we get
\begin{equation}
    \bra{n}D\ket{1}=\bra{n} \left[  e^{-\frac{S}{2}}\sum_{k=0}^\infty  \frac{S^{k/2}}{k!}\sqrt{(k+1)!}\ket{k+1} -\sqrt{S}D\ket{0}  \right].
\end{equation}
Using the orthonormality of the LHO eigenstates, we have
\begin{equation}
    \bra{n}D\ket{1}=e^{-\frac{S}{2}}\frac{S^{(n-1)/2}}{(n-1)!}\sqrt{n!} -\sqrt{S}\bra{n}D\ket{0}.
\end{equation}
Using Eq.~(\ref{eq:Dna0}), the first term is $\frac{n}{\sqrt{S}}\bra{n}D\ket{0}$, and we get the final result of the matrix element
\begin{equation}
    \bra{n}D\ket{1}=\left(\frac{n}{\sqrt{S}}-\sqrt{S}   \right)\bra{n}D\ket{0}.
\end{equation}

Combining the terms with Eq.~(\ref{eq:nQD0}), we arrive at:
\begin{equation}M = \mu_0 \langle n | D | 0 \rangle \left[ 1 + \frac{\mu'}{\mu_0} \sqrt{\frac{\hbar}{2M_{\text{eff}}\omega}} \left( \frac{n+S}{\sqrt{S}} \right) \right].
\end{equation}
Defining the dimensionless Herzberg-Teller coupling constant as $\alpha = \frac{\mu'}{\mu_0} \sqrt{\frac{\hbar}{2M_{\text{eff}}\omega}}$, the final expression for the vibronic emission intensity is:

\begin{equation}
I_{0\to n}^{\text{em}} \propto \frac{S^n}{n!} \left( 1 + \alpha \frac{n+S}{\sqrt{S}} \right)^2.\label{eq:HT_final}
\end{equation}
While the emission profile is governed by the matrix element $\langle n | Q D | 0 \rangle$, the absorption process requires the evaluation of the transition amplitude starting from the ground state and projecting onto the $n$-th vibrational level of the excited state manifold. This is represented by:
\begin{equation}M_{\text{abs}} = \langle D n | (\mu_0 + \mu' Q) | 0 \rangle = \mu_0 \langle n | D^\dagger | 0 \rangle + \mu' \langle n | D^\dagger Q | 0 \rangle.\end{equation}
In this case, the operator order in the second term is $D^\dagger Q$. Unlike the emission derivation in Eq.~(\ref{eq:nQD0}), where the commutator $[Q, D]$ introduced a $2\sqrt{S}$ term, the absorption matrix element depends directly on the action of the coordinate operator on the ground state. The vibronic correction then yields a factor proportional to $(n/\sqrt{S} - \sqrt{S})$. Consequently, the intensity distribution for absorption, including the sign change of $\alpha$ that stems from the $D^\dagger$ operator [$D^\dagger(\sqrt{S})=D(-\sqrt{S})$], is given by:
\begin{equation}I_{0\to n}^{\text{abs}} \propto \frac{S^n}{n!} \left( 1 + \alpha \frac{n-S}{\sqrt{S}} \right)^2.\end{equation}

This formulation accounts for the non-Condon effects observed in our experimental absorption spectra, where $\alpha$ parameterizes the vibronic borrowing strength relative to the purely electronic transition.

\bibliography{Bibliographyv2}

\end{document}